\documentclass[aps,pra,preprint,showpacs,groupedaddress]{revtex4}
\usepackage{graphicx}
\usepackage{amsmath}
\usepackage{longtable}

\begin{document}

\title{Polarizabilities of the Mg$^+$ and Si$^{3+}$ ions }
\author{J. Mitroy} 
\affiliation{School of Engineering, Charles Darwin University, Darwin NT 0909, Australia}  
\author{M. S. Safronova}  
\affiliation{Department of Physics and Astronomy, University of Delaware, Newark, Delaware 19716-2593, USA}

\date{\today}

\begin{abstract}  
A polarization analysis of the fine-structure intervals for the $n = 17$ 
Rydberg states of Mg and the $n = 29$ states of Si$^{2+}$ is performed.  
The coefficients of all terms in the polarization expansion  
up to $r^{-8}$ were computed using a semi-empirical single electron 
analysis combined with the relativistic all-order single-double method
(MBPT-SD) which includes all single-double excitations from the Dirac-Fock 
wave functions to all orders of perturbation theory.   
The revised analysis yields dipole polarizabilities of $\alpha_1 = 35.04(3)$ 
a.u. for Mg$^+$ and $\alpha_1 = 7.433(25)$ a.u. for Si$^{3+}$, values only 
marginally larger than those obtained in a previous analysis (E. L. Snow 
and S. R. Lundeen (2007) Phys.~Rev.~A {\bf 75} 062512, {\em ibid} (2008) 
{\bf 77} 052501).   The polarizabilities are used to make estimates of the 
multiplet strength for the resonant transition for both ions.  The  revised 
analysis did see significant changes in the slopes of the polarization plots.  
The dipole polarizabilities from the MBPT-SD calculation, namely 35.05(12) a.u. 
and 7.419(16) a.u., are within 0.3$\%$ of the revised experimental values. 

\end{abstract}  

\pacs{31.15.ap, 31.15.ag, 31.15.am, 31.15.V-}
\maketitle

\section{Introduction}

Resonant excitation Stark ionization spectroscopy (RESIS) \cite{lundeen05a} 
is a versatile and powerful method for studying Rydberg states of
atoms and ions.   One of the primary applications is the determination
of deviations from the pure hydrogenic values of the binding 
energies.  Polarization interactions between the core and the
Rydberg electrons lead to the effective potential 
\cite{drachman82b,drachman95a,lundeen05a} 
\begin{equation}
V_{\rm pol} = - \frac{C_4}{r^4} - \frac{C_6}{r^6} - \frac{C_7}{r^7} 
 - \frac{C_8}{r^8} - \frac{C_{8L}L(L+1)}{r^8} + \ldots  
\label{vpol} 
\end{equation}
This functional form has been applied to the analysis of the fine-structure 
spectrum of the Rydberg states of neutral Mg and Si$^{2+}$ resulting 
in precise estimates of the dipole polarizabilities of the sodium-like 
Mg$^+$ and Si$^{3+}$ ground states \cite{komara03a,snow07a,snow08a}. 
The Mg$^+$ polarizability was 35.00(5) a.u. \cite{snow08a} and the 
Si$^{3+}$ polarizability was 7.426(12) a.u. \cite{snow07a}.  Analysis 
of the spectrum has also given information about the quadrupole 
polarizabilities. 

One area of uncertainty in the analysis is the contribution of the
higher-order terms in the polarization expansion.  Using theoretical 
estimates of $C_7$ and $C_{8L}$ to constrain the analysis has proved 
essential in obtaining values of the quadrupole polarizability 
that are even remotely close with theoretical estimates 
\cite{snow07a,snow08a}.  However,  some of the high order terms that 
contribute to Eq.~(\ref{vpol}) were omitted from the analysis of the 
experimental data.    

This limitation is rectified in the present work which uses two 
different theoretical techniques to determine values of all the 
terms  in Eq.~(\ref{vpol}).  One technique supplements the
Hartree-Fock core potential with a semi-empirical polarization
potential and effectively solves a one-electron Schrodinger 
equation to determine the excitation spectrum for the valence electron 
\cite{muller84,mitroy88d,mitroy03g}.  The other method used is 
the relativistic all-order single-double 
method where all single and double excitations of the Dirac-Fock 
(DF) wave function are included to all orders of many-body 
perturbation theory (MBPT)  \cite{safronova98a,safronova99a,safronova08a}.  
We note in passing 
that there has been a recent configuration interaction (CI) calculation of 
the polarizabilities of the Mg$^{+}$ and Si$^{3+}$ ground states
\cite{hamonou07a}.

The current work has implications that go beyond the analysis 
of the RESIS experiments of the Lundeen group.  One of the
most active area in physics at present is the development 
of new atomic clocks based on groups of neutral atoms in optical 
lattices  \cite{gill05a,baillard08a,campbell08a} or 
single atomic ions \cite{diddams01a,gill05a}.  
These clocks have the potential to exceed the precision of the 
existing cesium microwave standard \cite{bauch03a}.  For many 
of these clocks the single largest source of systematic error 
is the black-body radiation shift (BBR) 
\cite{palchikov03a,porsev06b,zelevinsky07a,rosenband08a,itano07a}.  
The BBR shift to first order is proportional to the
difference in polarizabilities of the two states involved 
in the clock transition.  Many estimates of the relevant 
polarizabilities are determined by theoretical calculations
\cite{arora07a,mitroy08b,angstmann06a}.  Comparisons of 
existing techniques to calculate polarizabilities with high quality 
experiments will ultimately help constrain the uncertainties 
associated with the BBR shift. 

\section{The polarization expansion}

In this section the definitions of the various terms in the 
polarization potential are given following the analysis of 
Drachman \cite{drachman82b,drachman95a}.  The notation of Lundeen 
\cite{lundeen05a,snow07a} is adopted.    

The leading term, $C_4$ is half the size of the static dipole
polarizability,
\begin{equation}
C_4 = \frac{\alpha_1}{2} \ .  
\label{C4} 
\end{equation} 
The dipole polarizability is defined as 
\begin{equation}
\alpha_1 = \sum_n \frac{f^{(1)}_{gn}}{(\Delta E_{gn})^2} \ .  
\label{alphad} 
\end{equation} 
where $f^{(k)}_{gn}$ is the absorption oscillator strength 
for a dipole transition from state $g$ to state $n$.  The absorption 
oscillator strength for a multi-pole
transition from $g \to n$, with an energy difference
of $\Delta E_{ng} = E_{g} - E_{n}$, is defined as
\begin{equation}
f^{(k)}_{gn} =  \frac {2 |\langle \psi_{g};L_{g} \parallel  r^k
{\bf C}^{k}({\bf \hat{r}}) \parallel \psi_{n};L_{n} \rangle|^2 \Delta E_{ng}}
{(2k+1)(2L_g+1)}  \ .
\label{fvaldef}
\end{equation}
In this expression, $L_g$ is the orbital angular momentum of the
initial state while $k$ is the polarity of the transition.  In 
a $J$-representation, the oscillator strength becomes  
\begin{equation}
f^{(k)}_{gn} =  \frac {2 |\langle \psi_{g};J_{g} \parallel  r^k
{\bf C}^{k}({\bf \hat{r}}) \parallel \psi_{n};J_{n} \rangle|^2 \Delta E_{ng}}
{(2k+1)(2J_g+1)}  \ .
\label{fvaldefJ}
\end{equation}

The next term, $C_6$, is composed of two separate terms 
\begin{equation}
C_6 = \frac{\alpha_2 - 6\beta_1}{2}  \ .  
\label{C6} 
\end{equation}
The quadrupole polarizability, $\alpha_2$ is computed as 
\begin{equation}
\alpha_2 = \sum \frac{f^{(2)}_{gn}}{(\Delta E_{gn})^2} \ .  
\label{alphaq} 
\end{equation} 
The second term in Eq.~(\ref{C6}) is the non-adiabatic dipole
polarizability.  It is defined as  
\begin{equation}
\beta_1 = \sum \frac{f^{(1)}_{gn}}{2(\Delta E_{gn})^3} \ .  
\label{betad} 
\end{equation} 
The $r^{-7}$ term, $C_7$ also comes in two parts, namely
\begin{equation}
C_7 = -\frac{ (\alpha_{112} + 3.2 q \gamma_1 )}{2} .  
\label{C7} 
\end{equation}
The $\gamma_1$ is a higher-order non-adiabatic term  
\begin{equation}
\gamma_1 = \sum \frac{f^{(1)}_{gn}}{4(\Delta E_{gn})^4} \ .  
\label{gammad} 
\end{equation} 
while $q$ is the charge on the core.  The dipole-dipole-quadrupole polarizability, 
$\alpha_{112}$ arises from third order in perturbation theory.  It is derived 
from the matrix element  \cite{drachman82b,snow07a,snow07b} 
\begin{eqnarray}
\frac{\alpha_{112}}{2R^7} & = &  \sum_{k_1 k_2 k_3}  \sum_{n_a n_b}  
\frac{ \langle \psi_{g};0 |V^{k_1} | \psi_{n_a};L_{a} \rangle } {\Delta E_{n_g n_a} \Delta E_{n_b n_a} }  \nonumber  \\  
& \times &  \langle \psi_{n_a};L_{a} | V^{k_2} |  \psi_{n_b};L_{b} \rangle \langle \psi_{n_b};L_{b} | V^{k_3} | \psi_{g};0 \rangle \ .  
\label{alphaddq}
\end{eqnarray}
where $V^{k} = {\bf C}^{k}({\bf \hat{r}}) \cdot {\bf C}^{k}({\bf \hat{R}}) r^{k}/R^{K+1}$.    
The sum of the multipole orders must obey $k_1+k_2+k_3 = 4$.  
Quite a few terms contribute to $C_8$ 
\begin{equation}
C_8 = \frac{\alpha_3 - \beta_2 - \alpha_1 \beta_1 + \alpha_{1111} + 72\gamma_1}{2}  \ .  
\label{C8} 
\end{equation}
The octupole polarizability, $\alpha_3$ is computed as 
\begin{equation}
\alpha_3 = \sum \frac{f^{(3)}_{gn}}{(\Delta E_{gn})^2} \ .  
\label{alphao} 
\end{equation} 
The $\beta_2$ comes from the non-adiabatic part of the 
quadrupole polarizability, it is 
\begin{equation}
\beta_2 = \sum \frac{f^{(2)}_{gn}}{2(\Delta E_{gn})^3} \ .  
\label{betaq} 
\end{equation} 
The fourth-order term, $\alpha_{1111}$ is related to the hyper-polarizability
\cite{robb74a,king97a}.  It is defined as    
\begin{eqnarray}
\frac{\alpha_{1111}}{2R^8} &=&  \sum_{n_a n_b n_c}  
\frac{ \langle \psi_{n_g};0 |  V^1 | \psi_{n_a};L_{a} \rangle } 
{ \Delta E_{g a} \Delta E_{g b} \Delta E_{g c} } \ .  \nonumber \\  
& \times & \langle \psi_{n_a};L_{a} |  V^1 | \psi_{n_b};L_{b} \rangle \langle \psi_{n_b};L_{b} | V^1 | \psi_{n_c};L_{c} \rangle \nonumber \\  
& \times & \langle \psi_{n_c};L_{c} | V^1 | \psi_{g};0 \rangle \ .  
\label{alphadddd}
\end{eqnarray}
 
The final term, $C_{8L}$ is non-adiabatic in origin and defined  
\begin{equation}
C_{8L} = \frac{18 \gamma_1}{5} \ .   
\label{C8L} 
\end{equation}
\section{Structure models for M\lowercase{g}$^+$ and S\lowercase{i}$^{3+}$}

\subsection{Semi-empirical method}

The semi-empirical wave functions and transition operator expectation 
values were computed by diagonalizing the semi-empirical 
Hamiltonian \cite{mitroy03f,mitroy88d,mitroy93a,bromley02b,mitroy03e}
in a large mixed Laguerre type orbital (LTO) and Slater type
orbital (STO) basis set \cite{mitroy03f}.  We first discuss Si$^{3+}$ 
and then mention Mg$^+$.

The initial step was to perform a Hartree-Fock (HF) calculation to define
the core.  The present calculation can be regarded as HF plus core 
polarization (HFCP).  The calculation of the Si$^{3+}$ ground 
state was done in a STO basis \cite{mitroy99f}.  The core wave functions 
were then frozen, giving the working Hamiltonian for the 
valence electron 
\begin{eqnarray}
H  &=&  -\frac {1}{2} \nabla^2 + V_{\rm dir}({\bf r}) + V_{\rm exc}({\bf r})  
     +  V_{\rm p}({\bf r}) \ .  
\end{eqnarray}
The direct and exchange interactions, $V_{\rm dir}$ and $V_{\rm exc}$, of the 
valence electron with the HF core were calculated exactly.  The $\ell$-dependent 
polarization potential, $V_{\rm p}$, was semi-empirical in nature with the 
functional form
\begin{equation}
V_{\rm p}({\bf r})  =  -\sum_{\ell m} \frac{\alpha_d g_{\ell}^2(r)}{2 r^4}
                    |\ell m \rangle \langle \ell m| .
                                    \label{polar1}
\end{equation}
The coefficient, $\alpha_d$
is the static dipole polarizability of the core and 
$g_{\ell}^2(r) = 1-\exp\bigl(-r^6/\rho_{\ell}^6 \bigr)$
is a cutoff function designed to make the polarization potential 
finite at the origin.  The cutoff parameters, $\rho_{\ell}$ were 
tuned to reproduce the binding energies of the $ns$ 
ground state and the $np$, $nd$ and $nf$ excited states. 
The dipole polarizability for Si$^{4+}$ was chosen as  
$\alpha_d = 0.1624$ a.u. \cite{johnson83a,mitroy03f}.   
The cutoff parameters for $\ell = 0 \to 3$ were  
0.7473, 0.8200, 1.022 and 0.900 $a_0$ respectively.   
The parameters for $\ell > 3$ were set to $\rho_3$.   
The energies of the states with $\ell \ge 1$ were tuned to the 
statistical average of their respective spin-orbit doublets.  
The Hamiltonian was diagonalized in a very large orbital basis
with about 50 Laguerre type orbitals for each $\ell$-value.  
The oscillator strengths (and other multi-pole expectation values) 
were computed with operators that included polarization corrections 
\cite{hameed68a,hameed72a,vaeck92a,mitroy93a,mitroy03f}.  The 
quadrupole core polarizability was chosen as 0.1021 a.u. 
\cite{johnson83a} while the octupole polarizability was set to zero.   
The cutoff parameter for the polarization correction to the transition 
operator was fixed at 0.864 $a_0$ 
(the average of $\rho_0$, $\rho_1$, $\rho_2$ and $\rho_3$).

It is worth emphasizing that model potential is based on a realistic wave 
function and the direct and exchange interactions with the core were computed 
without approximation from the HF wave function.  Only the core polarization 
potential is described with an empirical potential.  

The overall methodology of the Mg$^+$ calculation is the same as that 
for Si$^{3+}$ and many of the details have been given previously 
\cite{mitroy08b}.  The core dipole polarizabilities were  
$\alpha_d = 0.4814$ a.u. \cite{muller84,mitroy03f} and 
$\alpha_q = 0.5183$ a.u. for Mg$^{2+}$  \cite{johnson83a,mitroy03f}. 
The octupole polarizability was set to zero.  
The Mg$^{2+}$ cutoff parameters for $\ell = 0 \to 3$ were 1.1795, 
1.302, 1.442, and 1.520 $a_0$ respectively.  The cutoff parameter for
evaluation of transition multipole matrix elements was 1.361 $a_0$.  

The HFCP calculations of the polarizabilities utilized the list of multipole 
matrix elements and energies resulting from the diagonalization of 
the effective Hamiltonian.  These were directly used in the evaluation 
of the polarizability sum rules.  

\subsection{The all-order method}

In the relativistic all-order method including single, double, and valence 
triple excitations, the wave function is represented as an expansion
\begin{eqnarray}
 |\Psi_v \rangle &=& \left[ 1 + \sum_{ma} \, \rho_{ma} \ a^\dagger_m a_a 
+ \frac{1}{2} \sum_{mnab} \rho_{mnab} \ a^\dagger_m
a^\dagger_n a_b a_a +
 \right.  \nonumber \\
& +& \left. \sum_{m \neq v} \rho_{mv} \ a^\dagger_m a_v + \sum_{mna}
\rho_{mnva} \ a^\dagger_m a^\dagger_n a_a a_v \right.
\nonumber \\
& + &\left. \frac{1}{6} \sum_{mnrab}
\rho_{mnrvab} \ a^\dagger_m a^\dagger_n a^\dagger_r a_b a_a a_v 
\right]|
\Phi_v\rangle, \label{eq1}
\end{eqnarray}
where $\Phi_v$ is the lowest-order atomic state wave function, which is
taken to be the {\em frozen-core} DF wave function of a 
state $v$ in our calculations.  In second quantization, 
the lowest-order atomic state function
is written as $$ |\Phi_v\rangle =a_v^{\dagger }|0_C\rangle,
$$ where $|0_C\rangle $ represents the DF wave function of the closed
core.  In Eq.~(\ref{eq1}), $a^\dagger_i$ and $a_i$ are
creation and annihilation operators, respectively.  Indices at the beginning of
the alphabet, $a$, $b$, $\cdots$,  refer to occupied core states, those in
the middle of the alphabet $m$, $n$, $\cdots$, refer to excited states,
and index $v$ designates the  valence orbital. We refer
to $\rho_{ma}$, $\rho_{mv}$ as single core and valence excitation 
coefficients and  to $\rho_{mnab}$ and $\rho_{mnva}$ as double core 
and valence excitation coefficients, respectively. The quantities $\rho_{mnrvab}$ 
are valence triple excitation coefficients and are included perturbatively where 
necessary as described in Ref.~\cite{safronova99a}.   
  
To derive the equations for the excitation coefficients, the wave 
function $\Psi_v$, given by Eq.~(\ref{eq1}), is substituted into 
the many-body Schr\"{o}dinger equation
\begin{equation}
H | \Psi_v\rangle=E| \Psi_v\rangle, \label{eq2}
\end{equation}
where the Hamiltonian $H$ is the relativistic {\em no-pair} 
Hamiltonian  \cite{brown51a}.  This can be expressed in second 
quantization as
\begin{equation}
H = \sum_{i} \epsilon_{i} :a_{i}^{\dagger} a_{i}:
+ \frac{1}{2} \sum_{ijkl} g_{ijkl} :a_{i}^{\dagger} a_{j}^{\dagger} a_{l} a_{k}: ,
\end{equation}
\noindent where  $\epsilon_{i}$ is the  DF energy for the 
state $i$, $g_{ijkl}$ are the two-body Coulomb integrals, and  
:~: indicates normal order of the operators
with respect to the closed core.
In the {\it no-pair } Hamiltonian, the contributions from
negative-energy (positron) states are omitted. 

The resulting all-order equations for the excitation coefficients $\rho_{ma}$, 
$\rho_{mv}$, $\rho_{mnab}$, and $\rho_{mnva}$ are solved iteratively with a 
finite basis set, and the correlation energy is used as a convergence parameter.
As a result, the series of correlation correction terms included in the SD (or SDpT) 
approach are included to all orders of many-body perturbation theory (MBPT) as 
an additional MBPT order is picked up at each iteration.  The basis set 
is defined in a spherical cavity on a non-linear grid and consists of 
single-particle basis states which are linear combinations of $B$-splines 
\cite{johnson88a}.  The contribution from the Breit interaction is negligible for 
all matrix elements considered in this work. 
  
The matrix element of any one-body operator $Z$ in the all-order method is 
obtained as
\begin{equation}
 Z_{vw} = \frac{\langle \Psi_v \vert Z \vert \Psi_w
\rangle } {\sqrt{\langle \Psi_v \vert \Psi_v \rangle \langle \Psi_w
\vert \Psi_w \rangle}}.
\end{equation}
The numerator of the resulting expression consists of the sum of
the DF matrix element $z_{wv}$ and twenty other terms $Z^{(k)}$,
$k=a\cdots t$. These terms  are
linear or quadratic functions of the excitation coefficients
$\rho_{ma}$, $\rho_{mv}$, $\rho_{mnab}$, and $\rho_{mnva}$.
More details on the SD and SDpT methods and their applications 
can be found in Refs.~\cite{blundell91a,safronova99a,safronova08a}.
We find that the contribution of triple excitations is small
for the atomic properties considered in this work. 
So the SD approximation is used for most transitions. 
 
The $B$-spline basis used in the calculations included  $N = 50$ basis orbitals 
for each angular momentum within a cavity radius of $R_0 = 100$ $a_0$ for Mg$^+$ 
and $R_0 = 80$ $a_0$ for Si$^{3+}$.  Such large cavities are needed to fit 
highly-excited states such as $8h$ needed for the $3d$ octupole polarizability 
calculations.  The single-double (SD) all-order method yielded results for the 
primary $ns-np_{j}$ electric-dipole matrix elements of alkali-metal atoms that 
are in agreement with experiment to 0.1\%-0.5\% \cite{safronova99a}.  We refer
to the results obtained with this method as MBPT-SD in the subsequent text and 
Tables.      


Since the all-order calculations are carried out with a finite basis set, the sums 
given by Eqs.~(\ref{alphad}) - (\ref{alphao}) run up to the number of the basis set 
orbitals ($N=50$) for each partial wave.  For consistency, the same $B$-spline 
basis is used in all calculations of the same system (e.g. Mg$^+$ or Si$^{3+}$).  

The calculation of the polarizabilities for the MBPT-SD uses slightly 
different procedures to include different parts of the polarizability 
sum rules.  The all-order matrix elements were combined with the 
experimental energies for excited states with $n\le6$ for 
$\beta = ns, np_{1/2}, np_{3/2}, nd_{3/2}, nd_{5/2}$, $n\le7$ for $\beta = nf_{5/2},
nf_{7/2}$, and $n\le8$ for  $\beta = ng_{7/2}, ng_{9/2}, nh_{9/2}, nh_{11/2}$.
The remaining matrix elements and energies were calculated in the DF approximation, with
the exception of the $3s$ dipole polarizability, where the remaining matrix elements were 
calculated
using random-phase approximation (RPA) \cite{johnson96a} for the purpose of error evaluation.
These remainder contributions are small for dipole polarizabilities (0.2-5\%)  but increase
in relative size for the quadrupole (0.3-10\%) and octupole (4-20\%) polarizabilities.
An extra correction was introduced to the remainder contribution for octupole 
polarizabilities.  First, the accuracy of the DF calculations was estimated from 
a comparison of the DF and all-order results for the few first terms.  Then, these
estimates were used to adjust the remainder.    The improvement of the DF results for 
states 
with higher $n$ was also taken into account. The size of this extra correction ranged 
from 0.9\% to 6\% of the tail contributions as the accuracy of the DF approximation 
for these highly-excited states is rather high.  The net effect of this scaling was
usually to reduce the octupole polarizabilities by an amount of about 0.5-1.5$\%$.  

The core contribution was calculated in the RPA \cite{johnson83a} with the 
exception of the dipole polarizability for the Mg$^{2+}$ core.  In this 
case  the polarizability of $\alpha_d = 0.4814$ a.u. was taken from 
a pseudo-natural orbital CI type calculation 
\cite{muller84,mitroy03f}. A 
small $\alpha_{cv}$ 
correction for the dipole polarizability that compensates for excitations from 
the core to occupied valence states was also determined using RPA 
matrix elements and DF energies.  The relative impact of the 
core polarizability was at least a factor of two smaller for 
the quadrupole polarizability.  

\begin{table}[th]
\caption[]{ 
Theoretical and experimental energy levels (in Hartree) of some
of the low-lying states of the Mg$^{+}$ and Si$^{3+}$ ions.  The 
energies are given relative to the energy of the 
Mg$^{2+}$ and Sr$^{4+}$ cores.  The experimental energies 
are taken from the National Institute of Standards and Technology 
database \cite{nistasd315}.  The HFCP energies should be interpreted 
as the $J$ weighted average of the spin-orbit doublet.  }
\label{elevels}
\begin{ruledtabular} 
\begin{tabular}{lccc}
  \multicolumn{1}{l}{State} &   Experiment  & MBPT-SD & HFCP \\ \hline
  \multicolumn{4}{c}{Mg$^+$}  \\ 
$3s_{1/2}$  & $-$0.552536   & $-$0.552522  & $-$0.552536    \\
$3p_{1/2}$  & $-$0.390015   & $-$0.390030  & $-$0.389737   \\
$3p_{3/2}$  & $-$0.389597   & $-$0.389611  &               \\
$4s_{1/2}$  & $-$0.234481   & $-$0.234470  & $-$0.234323   \\
$3d_{5/2}$  & $-$0.226803   & $-$0.226772  & $-$0.226804    \\
$3d_{3/2}$  & $-$0.226799   & $-$0.226768    \\
$4p_{1/2}$  & $-$0.185206   & $-$0.185210  & $-$0.185014     \\
$4p_{3/2}$  & $-$0.185067   & $-$0.185071   \\
$4d_{5/2}$  & $-$0.127382   & $-$0.127374  & $-$0.127373   \\
$4d_{3/2}$  & $-$0.127379   & $-$0.127372   \\
  \multicolumn{4}{c}{Si$^{3+}$}  \\ 
$3s_{1/2}$ & $-$1.658930  & $-$1.658973 &  $-$1.658928 \\
$3p_{1/2}$ & $-$1.334120  & $-$1.334094 &  $-$1.332738 \\
$3p_{3/2}$ & $-$1.332019  & $-$1.331999 &  \\
$3d_{5/2}$ & $-$0.928210  & $-$0.928138 &  $-$0.928302 \\
$3d_{3/2}$ & $-$0.928205  & $-$0.928134 \\
$4s_{1/2}$ & $-$0.775097  & $-$0.775104 &  $-$0.774681 \\
$4p_{1/2}$ & $-$0.664433  & $-$0.664421 &  $-$0.663640 \\
$4p_{3/2}$ & $-$0.663696  & $-$0.663684 \\
$4d_{5/2}$ & $-$0.519810  & $-$0.519800 &  $-$0.519743 \\
$4d_{3/2}$ & $-$0.519809  & $-$0.519799 \\
$4f_{5/2}$ & $-$0.501044  & $-$0.501044 &  $-$0.501033 \\
$4f_{7/2}$ & $-$0.501032  & $-$0.501035 \\
\end{tabular}
\end{ruledtabular} 
\end{table}

\section{Ground and excited properties}

\begin{table}[th]
\caption[]{ \label{Resfvals}
Line strengths (in a.u.) for the resonance transitions of Na, Mg$^+$,
Al$^{2+}$ and Si$^{3+}$.  Experimental values
with citations are also given.  The MBPT-SD results for Na and
Al$^{2+}$ are taken from \cite{safronova98a}.
 }
\begin{ruledtabular}
\begin{tabular}{lcccc}
Transition  & HFCP  &   \multicolumn{1}{c}{MBPT-SD}  &   BSR-CI  & Experiment  \\ \hline
  \multicolumn{5}{c}{Na}  \\
$S^{(1)}_{3s-3p_{1/2}}$  &  12.44    & 12.47  &  12.60 & 12.412(16) \cite{volz96a,volz96b} \\
                        &           &                  &         & 12.435(41) \cite{tiemann96a} \\
$S^{(1)}_{3s-3p_{3/2}}$  &  24.88    & 24.94  &  25.20 & 24.876(24) \cite{jones96a}  \\
                        &           &         &         & 24.818(34) \cite{volz96a,volz96b} \\
                        &           &         &         & 24.844(54) \cite{oates96a} \\
  \multicolumn{5}{c}{Mg$^+$}  \\
$S^{(1)}_{3s-3p_{1/2}}$  &  5.602    & 5.612     &  5.644  & 5.645(44) \cite{ansbacher89a}  \\
$S^{(1)}_{3s-3p_{3/2}}$  &  11.20    & 11.23    &  11.29  & 11.33(12) \cite{ansbacher89a}   \\
                         &           &          &         & 11.24(6) \cite{hermann08a}   \\
  \multicolumn{5}{c}{Al$^{2+}$}  \\
$S^{(1)}_{3s-3p_{1/2}}$  &  3.398    & 3.404   &  3.422  &  3.01(29) \cite{berry70a} \\
                         &          &          &         &  3.11(15) \cite{kernahan79a}     \\ 
                         &          &          &         &  3.31(35) \cite{baudinet79a}     \\ 
$S^{(1)}_{3s-3p_{3/2}}$  &  6.796    & 6.817   &  6.851  &  6.02(57) \cite{berry70a} \\ 
                         &          &          &         &  6.35(45)  \cite{kernahan79a}     \\ 
  \multicolumn{5}{c}{Si$^{3+}$}  \\
$S^{(1)}_{3s - 3p_{1/2}}$  &  2.333   & 2.341   &  2.350  & 2.35(10)  \cite{maniak93a} \\
$S^{(1)}_{3s - 3p_{3/2}}$  &  4.666   & 4.686   &  4.707  & 4.70(20)  \cite{maniak93a} \\
\end{tabular}
\end{ruledtabular}
\end{table}

\begin{table*}[th]
\squeezetable 
\caption[]{ \label{fvals}
Line strengths (in a.u.) for various transitions of Mg$^+$ and Si$^{3+}$.
The line strengths are mainly for dipole transitions with the
exception of the $3s \to 3d$ and $3s \to 4d$ transitions.  
 }
\begin{ruledtabular} 
\begin{tabular}{lcccccc}
  &  \multicolumn{3}{c}{Mg$^+$} &  \multicolumn{3}{c}{Si$^{3+}$}  \\
\cline{2-4}  \cline{5-7}  
Transition  & HFCP  &   \multicolumn{1}{c}{MBPT-SD}  &   BSR-CI  &  HFCP  &   \multicolumn{1}{c}{MBPT-SD}  &   BSR-CI    \\ \hline
$S^{(1)}_{3s-4p_{1/2}}$  &  0.00251  & 0.00261  &  0.00211  &   0.0382   &  0.0385  &  0.0385  \\
$S^{(1)}_{3s-4p_{3/2}}$  &  0.00501  & 0.00460  &  0.00362  &   0.0764   &  0.0744  &  0.0738     \\
$S^{(1)}_{3s-5p_{1/2}}$  &  0.00395  & 0.00402  &  0.00366  &   0.0138   &  0.0139  &  0.0138  \\
$S^{(1)}_{3s-5p_{3/2}}$  &  0.00790  & 0.00763  &  0.00692  &   0.0276   &  0.0270  &  0.0268     \\
$S^{(1)}_{3p_{1/2}-4s}$  &  2.887    &  2.868   &  2.886    &   0.6410   &  0.6334  &  0.6328    \\
$S^{(1)}_{3p_{3/2}-4s}$  &  5.773    &  5.779   &  5.815    &   1.282    &  1.284   &  1.283     \\
$S^{(1)}_{3p_{1/2}-5s}$  &           &  0.2117  &  0.2115   &            &  0.0633  &  0.0629    \\
$S^{(1)}_{3p_{3/2}-5s}$  &           &  0.4247  &  0.4243   &            &  0.1284   &  0.1267    \\
$S^{(1)}_{3p_{1/2}-3d_{3/2}}$  & 17.32   & 17.29   &  17.35 &     5.923   & 5.933    &  5.955   \\
$S^{(1)}_{3p_{3/2}-3d_{3/2}}$  & 3.463   & 3.468   &  3.482 &     1.185   & 1.190    &  1.195   \\
$S^{(1)}_{3p_{3/2}-3d_{5/2}}$  & 31.17   & 31.21   &  31.33 &     10.66   & 10.71    &  10.75   \\
$S^{(1)}_{3p_{1/2}-4d_{3/2}}$  & 0.4100  & 0.4168  &  0.4631 &    0.0234  & 0.0212   &  0.0204  \\
$S^{(1)}_{3p_{3/2}-4d_{3/2}}$  & 0.0820  & 0.0825  &  0.0918 &    0.00468 & 0.00455  &  0.00438  \\
$S^{(1)}_{3p_{3/2}-4d_{5/2}}$  & 0.7380  & 0.7423  &  0.8291 &    0.0421   & 0.0410   &  0.0395   \\
$S^{(2)}_{3s-3d_{3/2}}$  &  97.52   &  97.51   &             &    16.81    & 16.85  &     \\
$S^{(2)}_{3s-3d_{5/2}}$  &  146.3   & 146.3    &             &    25.21    & 25.27  &   \\
$S^{(2)}_{3s-4d_{3/2}}$  &  3.615   &  3.638   &             &    0.2732   & 0.2659 &     \\
$S^{(2)}_{3s-4d_{5/2}}$  &  5.422   &  5.455   &             &    0.4098   & 0.3986 &   \\
\end{tabular}
\end{ruledtabular} 
\end{table*}

\subsection{The energy levels}

The binding energies of the low-lying states of the Mg$^+$ and Si$^{3+}$ 
are tabulated and compared with experiment in Table \ref{elevels}.
The agreement between the HFCP energies and the experimental energies 
is generally of order 10$^{-4}$ Hartree. When the $\rho_{\ell}$
cutoff parameters are tuned to the lowest state of each symmetry
the tendency is for higher states of the same symmetry to be slightly 
under-bound.  The MBPT-SD binding energies generally agree with 
experiment to better than 10$^{-4}$ Hartree.  The MBPT-SD binding energies 
do not suffer any systematic tendency to either underbind or overbind 
as $n$ increases.     

\begin{table*}[th]
\squeezetable 
\caption[]{ \label{polarizabilities}
The polarizabilities for the $3s$, $3p$ and $3d$ states of Mg$^+$ 
and Si$^{3+}$.  The tensor polarizabilities are for the $M_J = J$ 
states.   For states with $\ell > 0$, the MBPT-SD  
average values represent the weighted values for the spin-orbit doublet.  }
\begin{ruledtabular}
\begin{tabular}{lcccccccc}
  \multicolumn{1}{l}{State}  & \multicolumn{2}{c}{$\alpha_1$ (au)} &   \multicolumn{2}{c}{$\alpha_{1,2JJ}$ (au)} &   \multicolumn{2}{c}{$\alpha_2$ (au)}  &  \multicolumn{2}{c}{$\alpha_3$ (au)}  \\
\cline{2-3} \cline{4-5} \cline{6-7} \cline{8-9}  
        &  HFCP  & MBPT-SD & HFCP  & MBPT & HFCP  & MBPT-SD &   HFCP & MBPT-SD  \\
\hline
Mg$^+$($3s$)           &  34.99  &  35.05    & 0         & 0        &  156.1   & 156.1  &  1715 &  1719 \\
Mg$^+$($3p_{1/2}$)     &         &  31.60    &    0      & 0     &       & 340.2     &        & 11778 \\
Mg$^+$($3p_{3/2}$)     &         &  31.88    & 1.162     & 1.156    &          & 343.0     &        & 11879 \\
Mg$^+$($3p$ - Average) &  31.79  &  31.79    &           &          &  341.7   &  342.1   &  11839 & 11845  \\
Mg$^+$($3d_{3/2}$)     &         &  189.3    & $-$78.47  & $-$79.15 &          & $-$9336  &   & 2.857$\times 10^5$  \\ 
Mg$^+$($3d_{3/2}$)     &         &  188.6    & $-$112.1  & $-$112.2 &          & $-$9341  &   & 2.860$\times 10^5$ \\ 
Mg$^+$($3d$ - Average) &  189.5  &  188.9    &           &          &  $-$9611 & $-$9339     &  2.855$\times 10^5$  &  2.859$\times 10^5$ \\ 
Si$^{3+}$($3s$)  &  7.399  &  7.419   & 0         & 0        &  12.13   & 12.15 &  47.03 & 47.15  \\
Si$^{3+}$($3p_{1/2}$)  &         &  3.120    &     0     &   0      &          &  13.05    &        & 155.1  \\
Si$^{3+}$($3p_{3/2}$)  &         &  3.183    &  1.459     & 1.462    &          &  13.21    &        & 157.1  \\
Si$^{3+}$($3p$ - Average)  &  3.158  &  3.162    &           &          &  13.17   &  13.16    &  156.3 & 156.5    \\
Si$^{3+}$($3d_{3/2}$)  &         &  5.168    & $-$0.6083  & $-$0.631 &          &  58.61    &        & 695.2  \\
Si$^{3+}$($3d_{5/2}$)  &         &  5.131    & $-$0.8690   & $-$0.848 &          &  58.61    &        & 696.2 \\
Si$^{3+}$($3d$ - Average)  &  5.135  &  5.146   &          &          &  58.43   & 58.61 &  693.2 & 695.8 \\
\end{tabular}
\end{ruledtabular}
\end{table*}

\begin{table}[th]
\caption[]{ \label{alphadetail}
Breakdown of the different contributions to the dipole polarizabilities of 
Mg$^+$ and Si$^{3+}$.  The $\epsilon p $ contribution includes 
both pseudo-state and continuum states.  Dipole polarizabilities from other 
sources are also listed with citation.  The estimated uncertainties 
for the different components of the uncertainty as estimated in brackets.
The RESIS reanalysis are taken from the reanalysis of the RESIS fine-structure
intervals described later.
}  
\begin{ruledtabular} 
\begin{tabular}{lcccc} 
  \multicolumn{1}{l}{Quantity} & \multicolumn{2}{c}{Mg$^+$} &  \multicolumn{2}{c}{Si$^{3+}$}    \\
                                \cline{2-3}   \cline{4-5}    
                               & \multicolumn{1}{c}{HFCP} &  \multicolumn{1}{c}{MBPT-SD}  & \multicolumn{1}{c}{HFCP} &  \multicolumn{1}{c}{MBPT-SD}    \\
\hline
$3s \to 3p$                       &  34.413   & 34.478(100)  &  7.153   & 7.180(6)  \\ 
$3s \to (4\!-\!6)p$               &  0.021    & 0.020(0)     &  0.054   & 0.053(0)  \\  
$3s \to \epsilon p$               &  0.091    & 0.087(4)     &  0.030   & 0.029(1)  \\  
Core                              &  0.481    & 0.481(10)    &  0.162   & 0.162(8)  \\  
Core-Valence                      &  $-$0.018 & $-$0.018(2)  & $-$0.005 & $-$0.005(1)  \\  
Total                             &  34.99    & 35.05(12)    &  7.394   & 7.419(16)  \\  
CI \cite{hamonou07a}              &  \multicolumn{2}{c}{35.66}     & \multicolumn{2}{c}{7.50}  \\  
RESIS \cite{snow07a,snow08a}      &  \multicolumn{2}{c}{35.00(5)}  & \multicolumn{2}{c}{7.426(12)}  \\  
Laser Exp \cite{lyons98a}         &  \multicolumn{2}{c}{33.80(50)} &                          \\  
$f$-sums  \cite{theodosiou95a}    &  \multicolumn{2}{c}{35.1}      &                          \\  
RCC  \cite{sahoo07a}              &  \multicolumn{2}{c}{35.04}     &                          \\  
RESIS reanalysis                  &  \multicolumn{2}{c}{35.04(3)}  & \multicolumn{2}{c}{7.433(25)}  \\  
\end{tabular}
\end{ruledtabular}
\end{table}

\begin{table}[th]
\caption[]{ \label{alpha2detail}
Breakdown of the different contributions to the quadrupole polarizabilities of 
Mg$^+$ and Si$^{3+}$.  The $\epsilon p $ contribution includes 
both pseudo-state and continuum states.  The quadrupole polarizabilities from a RESIS 
analysis is listed.   
 }
\begin{ruledtabular} 
\begin{tabular}{lcccc} 
  \multicolumn{1}{l}{Quantity} & \multicolumn{2}{c}{Mg$^+$} &  \multicolumn{2}{c}{Si$^{3+}$}    \\
                                \cline{2-3}   \cline{4-5}    
                               & \multicolumn{1}{c}{HFCP} &  \multicolumn{1}{c}{MBPT-SD}  & \multicolumn{1}{c}{HFCP} &  \multicolumn{1}{c}{MBPT-SD}    \\
\hline
$3s \to 3d$                     &  149.69  & 149.68(32) & 11.502   & 11.529(9)  \\ 
$3s \to (4\!-\!6)d$             &  4.99    & 5.01(4)    &  0.240   & 0.235(0)   \\  
$3s \to \epsilon d$             &  0.86    & 0.85(6)    &  0.289   & 0.280(8)   \\  
Core                            &  0.52    & 0.52(6)    &  0.102   & 0.102(12)   \\  
Total                           &  156.1   & 156.1(5)   &  12.13   & 12.15(3)   \\  
RESIS \cite{snow07a,snow08a}    &  \multicolumn{2}{c}{222(54)} &         \\  
RCC  \cite{sahoo07a}             &  \multicolumn{2}{c}{156.0}      &             \\  
\end{tabular}
\end{ruledtabular}
\end{table}

\begin{table}[th]
\caption[]{ \label{Cnvalues}
The polarizabilities and $C_n$ parameters computed from the composite 
list of HFCP and MBPT-SD matrix elements.  The parameters tabulated here 
can be regarded as the recommended theoretical values.  The $C_7$, $C_8$ 
and $C_{8L}$ parameters were used in the analysis the RESIS spectra for 
Mg$^+$ and Si$^{3+}$.   
 }
\begin{ruledtabular} 
\begin{tabular}{lcc} 
  \multicolumn{1}{l}{Quantity} & \multicolumn{1}{c}{Mg$^+$} &  \multicolumn{1}{c}{Si$^{3+}$}    \\
\hline
$\alpha_1$       &  35.05(12)     &   7.419(16) \\  
$\alpha_2$       &  156.1(5)      &  12.15(3)   \\ 
$\alpha_3$       &  1715(6)       &  47.03(12)   \\ 
$\beta_1$        &  106.0(3)      &  11.04(1)   \\ 
$\beta_2$        &  236.1(5)      &  8.065(6)   \\ 
$\gamma_1$       &  324.7(9)      &  16.82(1)   \\ 
$\alpha_{112}$   &  2416(52)      &  89.74(41)  \\ 
$\alpha_{1111}$  &  3511(90)      &  51.19(28)  \\ 
$C_4$            &  17.53(6)      &   3.710(8)  \\  
$C_6$            &  $-$240.1(12)  & $-$27.06(4) \\  
$C_7$            &  $-$1727(27)   & $-$125.6(2)  \\
$C_8$            &  10672(92)     &  553.1(6)    \\
$C_{8L}$         &  1169(3)       &  60.54(5)  \\
\end{tabular}
\end{ruledtabular} 
\end{table}

\subsection{Line strengths}

Table \ref{Resfvals} lists the line strengths for the resonant transitions  
of Na, Mg$^+$, Al$^{2+}$ and Si$^{3+}$.  All line strengths here and in 
the text below are given 
in a.u..  The HFCP values for sodium are 
from calculations previously reported in Ref. \cite{zhang07c} 
while the values for Al$^{2+}$ were taken from a calculation very similar 
in style and execution to the present calculations \cite{mitroy08m}. The 
MBPT-SD line strengths for Na and Al$^{2+}$ were taken from Ref. \cite{safronova98a}.  
Values from the extensive tabulation of dipole line strengths using a 
$B$-spline non-orthogonal configuration interaction with the Breit 
interaction (BSR-CI) \cite{fischer06a} are also listed.  The HFCP line 
strengths were computed from a common multiplet strength by multiplying
by the appropriate recoupling coefficients \cite{shore68a}.   

The comparisons for the resonant $3s$ $\to$ $3p$ transition reveal that 
the HFCP line strengths are the smallest, the BSR-CI line strengths are 
the largest and the MBPT-SD line strengths are intermediate between these 
two calculations.  The MBPT-SD dipole strengths are closer to HFCP for Na, 
Mg$^+$ and Al$^{2+}$ and about half-way between HFCP and BSR-CI for 
Si$^{3+}$.  The total variation between the three different calculations 
is about 1$\%$.  The most precise experiments performed on the Na-like  
iso-electronic series of atoms(ions) are those performed on sodium 
itself  \cite{volz96a,volz96b,tiemann96a,jones96a,oates96a}.  The 
experimental line strengths for sodium are in better agreement with 
the MBPT-SD and HFCP line strengths than they are with the BSR-CI line 
strengths.  

There have been two precision measurements of the $3s \to 3p$ 
transition rate for Mg$^+$.  The experiment of Ansbacher 
{\em et al.} \cite{ansbacher89a} gave slightly larger line 
strengths which agree best with the BSR-CI values.  However, the 
most recent trapped ion experiment \cite{hermann08a} gave a 
$3s_{1/2} \to 3p_{3/2}$ line strength of 11.24(6) that is  
in better agreement with the HFCP/MBPT-SD line strengths. 

Table \ref{fvals} lists the line strengths for a number of other 
dipole transitions for Mg$^+$ and Si$^{3+}$.  The line strengths for
the quadrupole $3s \to nd$ transitions are also listed due to their
importance in the determination of the quadrupole polarizabilities.

The $3p$ $\to$ $3d$ transition is the strongest transition emanating
from the $3p$ level.  The comparison between the three calculations exhibits 
a pattern similar to that of the resonant transition.  The HFCP line 
strengths are smallest, the BSR-CI line strengths are the largest, and 
the MBPT-SD line strengths lie somewhere between these two calculations.   

The astrophysically important Mg$^+$ $3s$ $\to$ $4p$ transition has a 
very small dipole strength.  It is close to the Cooper minimum 
\cite{fano68a} in the $3s$ $\to$ $np$ matrix elements and therefore is 
more sensitive to the slightly different energies between the spin-orbit 
doublet.  This caused the ratio of line strengths for the $4p_{1/2}$ and 
$4p_{3/2}$ transitions to deviate from the expected value of 2.  The MBPT-SD 
branching ratio of 1.76 agrees with the recent experimental values of 
1.74(6) \cite{sofia00a} and 1.82(8) \cite{fitzpatrick97a,sofia00a}.  
The HFCP multiplet strength of 0.00752 and the MBPT-SD multiplet strength  
of 0.00721 are about 5-10$\%$ smaller than the recent experimental estimates 
of 0.00793(26) \cite{sofia00a} and 0.00775(50) \cite{fitzpatrick97a,sofia00a}.  

There is also a deviation from the ratio of 2 for the 
$3s \to 4p_{1/2,3/2}$ transitions of Si$^{3+}$.  However, 
in this case the deviation is smaller.  Ratios of line strengths
for the stronger transitions are much closer to values expected
from purely angular recoupling considerations.  The $3p_{3/2}:3p_{1/2}$ 
ratio for Si$^{3+}$ was 2.002.   The $3p$ $\to$ $4s$ transition 
ratio has a slight deviation from 2, the MBPT-SD calculations giving 2.015 
for Mg$^+$ and 2.006 for Si$^{3+}$ (the BSR-CI ratios are similar).  

The better than 0.5$\%$ agreement between the model potential and MBPT-SD 
line strengths for strong transitions is consistent with previous comparisons.  
The general level of agreement between calculations with a semi-empirical 
core potential and more sophisticated ab-initio approaches for properties 
such as oscillator strengths, polarizabilities and dispersion coefficients 
has generally been very good \cite{mitroy03f,porsev06a,mitroy07d,mitroy07e}. 
There was a tendency for the agreement between the HFCP and MBPT-SD line
strengths to degrade slightly from Mg$^+$ and Si$^{3+}$.  This is 
probably due to the increased importance of relativistic effects 
as the nuclear charge increases.        

\subsection{Polarizabilities}

The polarizabilities of the $3s$, $3p$ and $3d$ levels of Mg$^+$ and Si$^{3+}$ 
are listed in Table \ref{polarizabilities}.  Tensor polarizabilities 
are also determined for the $3p$ and $3d$ levels.  Definitions of the 
tensor polarizability, $\alpha_{1,2JJ}$, in
terms of oscillator strength sum rules can be found in 
Refs.~\cite{safronova04a} and \cite{mitroy04b}.   

Table \ref{alphadetail} 
gives a short breakdown of the contributions of different terms to 
the dipole polarizability while Table \ref{alpha2detail} gives the
breakdown for the quadrupole polarizability.  The $3s \to \varepsilon p(d)$ 
contribution represents anything over $n = 6$ and can be regarded
as a mix of some higher discrete states as well as the pseudo-continuum.
Polarizabilities for 
the Mg$^+$ and Si$^{3+}$ ground states from other sources are also 
listed in Table \ref{alphadetail} and \ref{alpha2detail}.  The HFCP 
Mg$^+$ polarizability is marginally smaller than that reported previously 
\cite{mitroy08b} since the present evaluation includes a small core-valence
correction.    

The very good agreement between the HFCP and MBPT-SD polarizabilities 
is a notable feature of Table \ref{polarizabilities}.  
None of the static polarizabilities differ by more than 0.5$\%$ with 
the exception being the $\alpha_2$ of the Mg$^+$ $3d$ state.  Here
the difference is caused by the very small $\Delta E_{3d-4s}$ 
energy difference which is sensitive to small errors in 
the HFCP energies.   
The relative difference between some of the tensor polarizabilities 
is larger, but this is due to cancellations between the component 
sum rules that are combined to give the tensor polarizability.  

A recent CI calculation of the Mg$^+$ 
and Si$^{3+}$ ground state dipole polarizabilities \cite{hamonou07a} 
gave polarizabilities that were 1-2$\%$ larger than the HFCP/MBPT-SD 
polarizabilities.  The more recent relativistic coupled-cluster 
(RCC) calculation \cite{sahoo07a} gave polarizabilities that were 
compatible with the present values.   

The dipole polarizabilities 
for both Mg$^+$ and Si$^{3+}$ are dominated by the resonant oscillator 
strength.  For Mg$^+$ one finds that 98.3$\%$ of $\alpha_1$ arises from 
the $3s \to 3p$ transition. For Si$^{3+}$ the contribution is  
smaller but still substantial at 96.7$\%$.
The non-adiabatic dipole polarizabilities are even more dominated by the 
contribution from the resonant transition.  One finds that $99.9\%$ of 
$\beta_1$ and 99.99$\%$ of $\gamma_1$ for Mg$^+$ come from this transition.
The proportions for the Si$^{3+}$ $\beta_1$ and $\gamma_1$ are 99.6$\%$ 
and $99.92\%$ respectively.

The quadrupole polarizabilities are also dominated by a single 
transition.  Table \ref{alpha2detail} shows that the $3s \to 3d$ 
excitation constitutes at least 95$\%$ of $\alpha_2$ for both Mg$^+$ 
and Si$^{3+}$.   

The calculation of the $\alpha_{112}$ and $\alpha_{1111}$ polarizabilities 
was a composite calculation using both MBPT-SD and HFCP matrix elements.  The 
HFCP calculation automatically generates a file containing matrix elements 
between every state included in the basis.  The more computationally 
intensive MBPT-SD calculation was used for the largest and most important 
matrix elements.  The HFCP matrix elements for the $3s \to 3p$, 
$3p \to 3d$, $3s \to 3d$ and $3p \to 4s$ transitions were replaced 
by $J$ weighted averages of the equivalent MBPT-SD matrix elements.  
This procedure combines the higher accuracy of the MBPT-SD calculation 
with the computational convenience of the HFCP calculation.  The 
justification for this procedure is that the predominant contribution 
to the polarizability comes from the low-lying transitions.   The 
resulting polarizabilities are listed in Table \ref{Cnvalues}.  The 
biggest change in $\alpha_{112}$ and $\alpha_{1111}$ resulting  
from using the composite matrix element list was less than $0.3\%$.
The $\alpha_{112}$ and $\alpha_{1111}$ polarizabilities did not 
allow for contributions from the core.  The impact of the core will 
be small due to large energy difference involving core excitations.  
The relative effect of the core for $\alpha_{112}$ and $\alpha_{1111}$ 
can be expected to be about as large as the core effect in the ground 
state $\alpha_1$ and $\alpha_2$ since there are core excitations that 
contribute to with only one core energy in the energy denominator.  
For example, consider the $\alpha_{112}$ excitation sequence of 
$2p^6 3s$ $^2$S$^e$ $\to$ $2p^5 3s 3d$ $^2$P$^o$ $\to$  $2p^6 3d$ 
$^2$D$^e$ $\to$ $2p^6 3s$ $^2$S$^e$.
The numerical procedures used to generate the 
$\alpha_{112}$ and $\alpha_{1111}$ polarizabilities were validated 
for He$^+$.  A calculation of the He$^+$ excitation spectrum 
was performed and the resulting lists of reduced matrix elements were entered
into the polarizability programs.  All the coefficients given by
Drachman \cite{drachman82a} were reproduced.       

The polarizabilities in Table \ref{Cnvalues} from the composite matrix
element list could be regarded as the recommended set of polarizabilities.  
The MBPT-SD matrix elements are used for the dominant low-lying transitions.  
The HFCP matrix elements are more accurate than the RPA/DF matrix 
elements used for the $3s \to \varepsilon p(d)$ remainders.  The only 
difference between the Table \ref{Cnvalues} and MBPT-SD 
polarizabilities occurs for $\alpha_3$.  A relatively large part of 
of $\alpha_3$ comes from the higher excited states and the continuum.  
Accumulating a lot of small contributions is tedious for the 
computationally expensive MBPT-SD, so this is done with the less
accurate DF approach.  In this case the HFCP polarizability is 
to be preferred.  It should be noted that the octupole polarizability 
is of minor importance in the subsequent analysis.     

\subsection{Error assessment}

Making an a-priori assessment of the accuracy of the HFCP 
polarizabilities is problematic since they are semi-empirical 
in nature.  The error assessment for the MBPT-SD proceeds by 
assuming that the total contribution of fourth- and higher-order 
terms omitted by the SD all-order method does  
not exceed the contribution of already included fourth- 
and higher-order terms.  Thus, the uncertainty of the SD 
matrix elements is estimated to be the difference between
the SD all-order calculations and third-order results.     

This procedure was applied to the $S_{3s-3p_{1/2}}$ line strength of 
sodium yielding an uncertainty $\delta S_{3s-3p_{1/2}}=0.092$.  
This uncertainty exceeds the difference 
between the SD line strength of 12.47 \cite{safronova98a} and recent 
high precision experiments which give 12.412(16) \cite{volz96a,volz96b}, 
and 12.435(41) \cite{tiemann96a}.  A similar situation applies for
the $S_{3s-3p_{3/2}}$ line strength.  

A detailed first principles evaluation of the uncertainty of the Si$^{3+}$ 
static dipole polarizability has been done and the uncertainty budget 
is itemized in Table \ref{alphadetail}.  In this case, the difference 
between the SD line strength and third order line strength for the
resonance transition was 0.082$\%$.  The uncertainties in the remaining 
($n = 4 - 6$) discrete transitions were of similar size.  
Uncertainties in the energies used in the oscillator strength sum rule can 
be regarded as insignificant since experimental energies were used.  
To estimate the accuracy of the remainder of the valence sum, the 
($n = 4 - 6$) calculation was repeated using RPA matrix elements and 
DF energies.  The difference of $3\%$ between the MBPT-SD and DF/RPA  
values was assessed to be the uncertainty in the $\varepsilon p$ remainder.  
The good agreement between the HFCP and DF/RPA for the non-resonant 
valence contribution gives additional evidence that the uncertainty 
estimate is realistic. 

The core dipole polarizability calculated in the RPA is known to 
underestimate the actual core polarizability.  For neon, the RPA 
gives $\alpha_1 = 2.38$ a.u. \cite{johnson83a} which is 11$\%$ smaller 
than the experimental value of 2.669 a.u. \cite{kumar85b}.  For Na$^+$,
the RPA gives 0.9457 a.u. \cite{johnson83a} while experiment gives 
1.0015(15) a.u. \cite{freeman76a}.   The pseudo-natural orbital approach 
used for Mg$^{2+}$ gave $\alpha_1 = 2.67$ a.u. for Ne \cite{werner76a} 
and $\alpha_1 = 0.9947$ a.u. for Na$^+$ \cite{muller84}.  
The uncertainty in the quadrupole core polarizability is based on 
comparisons with coupled cluster calculations for neon 
\cite{maroulis89a,nicklass95a}.  The RPA value of 6.423 a.u. is
about 12$\%$ smaller than the coupled cluster values of 7.525 a.u. 
\cite{nicklass95a} and 7.525 a.u. \cite{maroulis89a}.  The relative 
uncertainties are $\delta \alpha_1(\text{Mg}^{2+}) = 2$$\%$,      
$\delta \alpha_1(\text{Si}^{4+}) = 5$$\%$,      
$\delta \alpha_2(\text{Mg}^{2+}) = 12$$\%$, and       
$\delta \alpha_2(\text{Si}^{4+}) = 12$$\%$.       
The core-valence 
correction was assigned an uncertainty of 20$\%$ based on differences
between DF and RPA matrix elements.  The RPA error estimates are 
likely to be very conservative since the uncertainty in the RPA   
polarizabilities is expected to decrease as the nuclear charge 
increases. 

Combining the uncertainties in the valence and core polarizabilities 
for Si$^{3+}$ gives a final uncertainty of 0.16 a.u. (or $0.22\%$) 
in the MBPT-SD $\alpha_1$.

The uncertainty in the Si$^{3+}$ $\alpha_2$ listed in Table 
\ref{alpha2detail} was evaluated with a process  
that was similar to the dipole polarizability.  The difference 
between the SD line strength and third order line strength for the
$3s \! \to \! 3d_{5/2}$ transition was 0.064$\%$ (the relative 
uncertainty was almost the same for the transition to the 
$3d_{3/2}$ state).  This uncertainty is slightly smaller than that 
for the resonant dipole transition.  This was expected since the 
$3d$ electron is further away from the nucleus than the 
$3p$ electron and therefore correlation-polarization corrections 
have less importance.  Rather than do a computationally expensive
analysis, the relative uncertainties in the ($nd$ + $\epsilon d$)  
remainders were conservatively assigned to be same as for the
dipole transitions.  The final uncertainty was 
$\delta \alpha_2 = 0.03$ a.u.. 

The relative uncertainties in the Mg$^+$ polarizabilities are set 
in the same way as Si$^{3+}$.  The difference between the third-order 
and all-order dipole line strengths for the resonance transition was 
0.3$\%$.  The relative differences were larger for the $n = 4-6$ 
transitions due to their small size.  For example, the third-order/all-order 
comparison for the $S_{3s - \! 4p}$ multiplet strength gave 5$\%$.  This 
is consistent with the difference with the experimental multiplet 
multiplet strength.  The uncertainties were slightly smaller for 
the slightly larger $5p$ and $6p$ transitions.  However, the net
contribution to the uncertainty was miniscule since the line 
strengths were so small.
The $3s \to \varepsilon p$ uncertainty of 5$\%$ was based on
differences between the HFCP and DF/RPA matrix elements.

The uncertainties in the Mg$^+$ $\alpha_2$ polarizability are listed 
in Table \ref{alpha2detail} and $n = 3-6$ transitions were derived 
from the third-order/all-order comparison.  The relative uncertainty 
in the $3s \to 3d$ transition was 0.22$\%$.  
The very good agreement between the HFCP and MBPT-SD values for 
these terms is further supportive of a small uncertainty for
the $n = 3-6$ transitions. The 7$\%$ uncertainty in the 
$3d \to \varepsilon d$ remainder was based on the differences between 
the MBPT-SD and DF matrix elements.   

The relative uncertainties in the octupole polarizabilities listed 
in Table \ref{Cnvalues} were set to the uncertainties in the
quadrupole polarizabilities.  The $nf$ orbitals are further
away from the core than the $3d$ orbitals and so the $\alpha_2$ 
uncertainty serves as a convenient overestimate.  

The uncertainties in the higher-order polarizabilities $\beta_1$,
$\beta_2$ and $\gamma_1$ listed in Table \ref{Cnvalues} were taken to 
be the uncertainties in the resonant line strengths.  The higher powers 
in the energy denominator means other transition make a negligible 
contribution.  

The uncertainties in $\alpha_{112}$ and $\alpha_{1111}$  
were derived from the uncertainties in the reduced matrix elements.   
The relative uncertainties for the most important 
$3s \to 3p$, $3p \to 3d$ and $3s \to 3d$ matrix elements were simply 
added to give relative uncertainties for valence part of $\alpha_{112}$ 
and $\alpha_{1111}$.  The relative uncertainty resulting from the 
omission of core excitations was taken as the ratio of the core
to total dipole polarizability and added to the $\alpha_{112}$ and 
$\alpha_{1111}$ uncertainties.      

The uncertainties in $C_6$, $C_7$, $C_8$ and $C_{8L}$ were determined 
by combining the uncertainties of the constituent 
polarizabilities.  The most important of these parameters is the 
expected slope of the polarization plot, i.e.  
$\delta C_6 = \delta \alpha_2/2 + 3 \delta \beta_1$.    
For Si$^{3+}$ we get $\delta C_6 = 0.015+0.027=0.042$.   
For Mg$^+$ the uncertainty was $\delta C_6 = 1.2$.

\section{Polarization analysis of Rydberg States}

\subsection{The polarization interaction}

The various polarizabilities needed for the polarization analysis 
are listed in Table \ref{Cnvalues}.  The $C_7$, $C_8$ and $C_{8L}$ 
values were used to make corrections to the experimental energy 
intervals.  The $C_4$ value was used in computing the second-order energy 
shift.  The transition matrix elements used in this calculation 
represent a synthesis of the HFCP and MBPT-SD calculations.  

Estimates of $C_7$ and $C_{8L}$ were previously made by Snow and 
Lundeen \cite{snow07a} using MBPT-SD transition amplitudes for the 
lowest lying transitions.  These earlier estimates are within a
few percent of the present more sophisticated analysis.  The Snow 
and Lundeen values for $C_7$ were $-$1684(9) a.u. for Mg$^+$ and
$-$122(9) a.u. for Si$^{3+}$.  They are a few percent smaller 
than those listed in Table \ref{Cnvalues} due 
to the omission of higher excitations from the sum rule. The    
Snow and Lundeen values for $C_{8L}$ were 1170(12) a.u. for Mg$^+$ 
and 60.5 a.u. for Si$^{3+}$.  

One aspect of Table \ref{Cnvalues} that is relevant to the interpretation 
of experiments is the importance of the non-adiabatic dipole 
polarizabilities.  Consider Mg$^+$ for example.  The respective 
contributions to $C_6$ are 78.05 a.u. from $\alpha_2$ and $-318.0$ a.u. 
from  $-6\beta_2$.  Similarly, one finds that the $\gamma_1$ term of 
$-1.6\times 324.7$ makes up 30$\%$ of the final $C_7$ value of $-$1727 a.u.  
And finally, one finds that the $C_8$ value of 10672 is largely
due to the $36 \gamma_1$ contribution of 11689 a.u..  The degree
of importance of the non-adiabatic terms scarcely diminishes for
the Si$^{3+}$ ion. 

Table \ref{DeltaE} gives the energy shifts to the $n=17$ levels of 
Mg$^+$ and the $n=29$ levels of Si$^{3+}$ using the values in Table 
\ref{Cnvalues}.  The energy shifts need $\langle r^{-n} \rangle$ 
expectation values which were evaluated using the formulae 
of Bockasten \cite{bockasten74a}.    

\subsection{The polarization plot}

Polarizabilities can be extracted from experimental data by using a 
polarization plot.  This is based on a similar procedure that is 
used to determine the ionization limits of atoms \cite{edlen64a}.  
The notations $B_4$ and $B_6$, (instead of $C_4$ and $C_6$) are used 
to represent the polarization 
parameters extracted from the polarization plot.  This is to clearly 
distinguish them from polarization parameters coming from atomic 
structure calculation.
Assuming the dominant terms leading to departures from hydrogenic 
energies are the $B_4$ and $B_6$ terms, one can write 
\begin{equation}
\frac{\Delta E}{\Delta \langle r^{-4} \rangle} = 
B_4 + B_6 \frac{\Delta \langle r^{-6} \rangle}{\Delta \langle r^{-4} \rangle} \ .  
\label{edlen1} 
\end{equation} 
In this expression, $\Delta E$ is the energy difference between two states of 
the same $n$ but different $L$, while 
$\Delta \langle r^{-6} \rangle$ and $\Delta \langle r^{-4} \rangle$ 
are simply the differences in the radial expectations of the two
states.    

There are other corrections that can result in Eq.~{(\ref{edlen1}) 
departing from a purely linear form.  These are relativistic
energy shifts, Stark shifts due to a residual electric field, and
polarization shifts due to the $C_7$, $C_8$ (and possibly higher-order) 
terms of Eq.~(\ref{vpol}).  The energy difference between the   
$(n,L)$ and $(n,L')$ states can be written 
\begin{eqnarray}
\Delta E & = & \Delta E_4 + \Delta E_6 + \Delta E_7 + \Delta E_8 + \Delta E_{8L}  \nonumber \\ 
         & + & \Delta E_{\rm rel} + \Delta E_{\rm sec} + \Delta E_{\rm ss} \ , 
\label{DeltaE1} 
\end{eqnarray} 
where $\Delta E_n$ arises from the polarization terms of order 
$\langle r^{-n} \rangle$.   

Dividing through by $\Delta \langle r^{-4} \rangle$ and replacing
$\Delta E_6$ by $B_6 \Delta \langle r^{-6} \rangle$ gives   
\begin{eqnarray}
\frac{\Delta E}{\Delta \langle r^{-4} \rangle} &=& B_4 + 
B_6 \frac{\Delta \langle r^{-6} \rangle} {\Delta \langle r^{-4} \rangle}  
+  \frac{\Delta E_7 + \Delta E_8+ \Delta E_{8L}}{\Delta \langle r^{-4} \rangle} \nonumber \\  
&+&  \frac{\Delta E_{\rm rel} + \Delta E_{\rm sec} + \Delta E_{\rm ss}} {\Delta \langle r^{-4} \rangle} \ .  
\label{DeltaE2} 
\end{eqnarray} 

The influence of the Stark shifts, relativistic shifts, and second-order
polarization correction can be incorporated into the polarization plot 
by simply subtracting the energy shifts.  The corrected energy shift,
$\Delta E_{\rm c1}$, is defined as  
\begin{eqnarray}
\frac{\Delta E_{c1}}{\Delta \langle r^{-4} \rangle} = 
  \frac{\Delta E_{\rm obs}}{\Delta \langle r^{-4} \rangle}   
-  \frac{\Delta E_{\rm rel} + \Delta E_{\rm sec} + \Delta E_{ss}} {\Delta \langle r^{-4} \rangle} \ . 
\label{DeltaEc1} 
\end{eqnarray} 
An approximate expression is used for the relativistic energy 
correction.  This is taken from the result  
\begin{equation}
E_{\rm rel} = - \frac{\alpha^2 Z^4}{2n^3}\left( \frac{1}{j+1/2} - \frac{3}{4n} \right) \ . 
\label{Erel} 
\end{equation} 
The correction due to second-order effects, $\Delta E_{\rm sec}$,  uses the 
results of Drake and Swainson \cite{drake91a}.  The Stark shift corrections
use the Stark shift rates from Snow and Lundeen \cite{snow07a,snow08a} and
the deduced electric field.   
The energy corrections due to relativistic and polarization effects for
the states of Mg$^+$ and the Si$^{3+}$ for which RESIS data existed are 
listed in Table \ref{DeltaE}.

The second corrected energy is defined by further subtracting
the polarization shifts, $\Delta E_7$, $\Delta E_8$ and $\Delta E_{8L}$, 
\begin{equation}
\frac{\Delta E_{c2}}{\Delta \langle r^{-4} \rangle} = 
  \frac{\Delta E_{\rm c1}}{\Delta \langle r^{-4} \rangle}  
-  \frac{\Delta E_7 + \Delta E_8 + \Delta E_{8L}}{\Delta \langle r^{-4} \rangle} \ .  
\label{DeltaEc2} 
\end{equation} 

\subsection{Mg$^{+}$}

The energy splitting between adjacent $L$ Rydberg levels is dominated 
by the $C_4$ term.  The next biggest term is the $\Delta E_6$ term which 
is 3$\%$ of $\Delta E_4$ for the (17,6)-(17,7) interval.  The 
$\Delta E_{8L}$ correction is larger than $\Delta E_7$. The relative 
impact of the higher-order corrections diminishes as $L$ increases. 

The revised analysis of the RESIS energy intervals for Mg$^+$ was performed 
by subtracting the $\Delta E_{\rm c1}$ and $\Delta E_{\rm c2}$ energy 
corrections itemized in Table \ref{DeltaE} from the observed  
energy splittings.    This represents a refinement over the previous 
analysis by Snow and Lundeen \cite{snow08a} in a couple of respects.  
First, Snow and Lundeen did not include the $C_8$ term since the 
necessary polarizability information simply was not available.   
Their evaluation of $\alpha_{112}$ only included the $3p$ and $3d$ 
states in the intermediate sums.  The 
truncation of the sums in the $\alpha_{112}$ calculation was justified 
as the correction to $\alpha_{112}$ from a more complete evaluation 
was only a few percent.  The impact of the $\Delta E_8$ shift is more 
substantial.  Table \ref{DeltaE} shows the relative size of 
$\Delta E_8$ with respect to $\Delta E_7$ ranging from 35$\%$ to 11$\%$. 

Figures \ref{mg+} shows the polarization plot for Mg$^+$. Linear
regression was applied to the four data points with  
$\Delta \langle r^{-6} \rangle /\Delta \langle r^{-4} \rangle < 0.002$.
The (17,6)-(17,7) interval was omitted from the fit because the 
influence of $\Delta E_{7,8,8L}$ and $\Delta E_{\rm sec}$ amount 
to just over 50$\%$ of $\Delta E_6$.  Visual examination 
of Figure \ref{mg+} shows this data point lies a significant distance 
away from the line of 
best fit obtained from the four remaining points.  The linear
regression gave an intercept of $B_4 = 17.522(7)$ a.u. and a slope of 
$B_6 = -251.2(79)$ a.u..  The quoted uncertainties  are the statistical
uncertainties from the linear regression fit.    
 
The new value of the dipole polarizability derived from the 
polarization plot intercept was 35.044 a.u..  This is marginally 
larger than the polarizability of 35.00(5) a.u. given in the original 
Snow and Lundeen analysis \cite{snow08a}.  The present $\alpha_1$ 
is larger because the additional corrections in the 
$\Delta E_{\rm c2}$ energies lead to a steeper polarization plot.    

The slope of $B_6 = -251.2(79)$ is slightly steeper than the Table 
\ref{Cnvalues} recommended 
$C_6$ of $-$240.1(12).  Using the slope of $-$251.2 in conjunction with 
a $\beta_1 = 106.0$ a.u. gives a quadrupole polarizability of 
$\alpha_2 = -502.4 + 636.3 = 133.9$ a.u..  This is 
about 90 a.u. smaller than the polarizability of 222(54) a.u..  
given by Snow and Lundeen \cite{snow08a}.  However it is only 
22 a.u. smaller than the theoretical polarizabilities of 156.1 a.u..  
The uncertainty in the derived quadrupole polarizability would
be $(2 \times 7.9+ 6 \times 0.3) = 17.6$.  The RESIS and theoretical 
values are slightly outside their respective combined error 
estimates.  However, the uncertainty 
estimate used for $C_6$ is purely statistical in nature and does
not allow higher order corrections to Eq.~(\ref{edlen1}). 
This point is discussed in more detail later.      

The relatively large change in $\alpha_2$ from 222(54) to 134(18) a.u. was 
caused by the inclusion of $\Delta E_8$.  There is a near cancellation 
between some of the $\Delta E_7$ and $\Delta E_{8L}$ energy corrections.  
Hence the inclusion of the $\Delta E_8$ energy correction has a relatively 
large impact.  For example, the sum of $\Delta E_7$ and $\Delta E_{8L}$ 
for the (17,7)-(17,8) interval was 1.473 MHz.  The $\Delta E_8$ 
correction was 0.851 MHz. 

The derived dipole polarizability and value of $B_6$ are not sensitive 
to small changes in the $C_n$ values used 
for the corrections.  An analysis using alternate $C_n$ values derived 
from the uncertainties detailed in Table \ref{Cnvalues} was performed.  
This resulted in an additional uncertainty of 0.0004 a.u. in $B_4$ and 
an additional uncertainty of 1.6 in $B_6$.  These additional 
uncertainties were sufficiently small to ignore in subsequent 
analysis. 

\begin{table*}[tbp]
\squeezetable 
\caption[]{ \label{DeltaE}
Various energy corrections (in units of MHz) for the $n = 17$ intervals of Mg$^+$ and 
the $n = 29$ intervals of Si$^{3+}$.  These were computed using $C_n$ values of Table 
\ref{Cnvalues}.    
  }  
\begin{ruledtabular} 
\begin{tabular}{lllcccccccc}
 \multicolumn{1}{l}{$n$} & \multicolumn{1}{l}{$L_1$}  & $L_2$ &  \multicolumn{1}{c}{$\Delta E_{\rm rel}$ }  &  \multicolumn{1}{c}{$\Delta E_4$ }  
   & \multicolumn{1}{c}{$\Delta E_{\rm 6}$ }  &  \multicolumn{1}{c}{$\Delta E_7$ }  &  \multicolumn{1}{c}{$\Delta E_8$ }    
   & \multicolumn{1}{c}{$\Delta E_{\rm 8L}$ }  & \multicolumn{1}{c}{$\Delta E_{\rm sec}$ } & \multicolumn{1}{c}{$\Delta E_{\rm ss}$}  \\ \hline  
\hline
    \multicolumn{11}{c}{Mg$^{+}$  }   \\ 
 17 &  6 &   7 &     0.7314 &  1555.7935 &  $-$53.7751 &  $-$20.4880 &   7.1989 &   31.5551 &   8.1506 &  $-$0.1122 \\
 17 &  7 &   8 &     0.5593 &   678.8962 &  $-$12.6733 &   $-$3.4612 &   0.8512 &    4.9308 &   1.5012 &  $-$0.1702 \\
 17 &  8 &   9 &     0.4416 &   326.8907 &   $-$3.5557 &   $-$0.7292 &   0.1327 &    0.9816 &   0.3393 &  $-$0.2320 \\
 17 &  9 &  10 &     0.3575 &   169.8765 &   $-$1.1373 &   $-$0.1808 &   0.0253 &    0.2325 &   0.0896 &  $-$0.2723 \\
 17 & 10 &  11 &     0.2953 &    93.8138 &   $-$0.4026 &   $-$0.0508 &   0.0056 &    0.0627 &   0.0267 &  $-$0.3039 \\
    \multicolumn{11}{c}{Si$^{3+}$ }    \\ 
 29 &  8 &   9 &     7.2052 &  1172.2322 &  $-$67.1286 &  $-$27.9320 &  11.3800 &   83.6138 &   2.4220 &  $-$0.1658 \\
 29 &  9 &  10 &     5.8328 &   614.9317 &  $-$22.2670 &   $-$7.2876 &   2.3124 &   21.0913 &   0.6574 &  $-$0.3049 \\
 29 & 10 &  11 &     4.8184 &   343.2123 &   $-$8.2257 &   $-$2.1728 &   0.5526 &    6.1258 &   0.2026 &  $-$0.4199 \\
 29 & 11 &  12 &     4.0474 &   201.5328 &   $-$3.3152 &   $-$0.7211 &   0.1502 &    1.9896 &   0.0693 &  $-$0.6026 \\
 29 & 11 &  13 &     7.4953 &   324.9889 &   $-$4.7508 &   $-$0.9824 &   0.1956 &    2.6972 &   0.0951 &  $-$1.3540 \\
 29 & 11 &  14 &    10.4675 &   403.3750 &   $-$5.4110 &   $-$1.0843 &   0.2106 &    2.9685 &   0.1054 &  $-$2.3123 \\
\end{tabular}
\end{ruledtabular} 
\end{table*}
                                                                                                                  
\begin{figure} [th]
\includegraphics[width=8.50cm,angle=0]{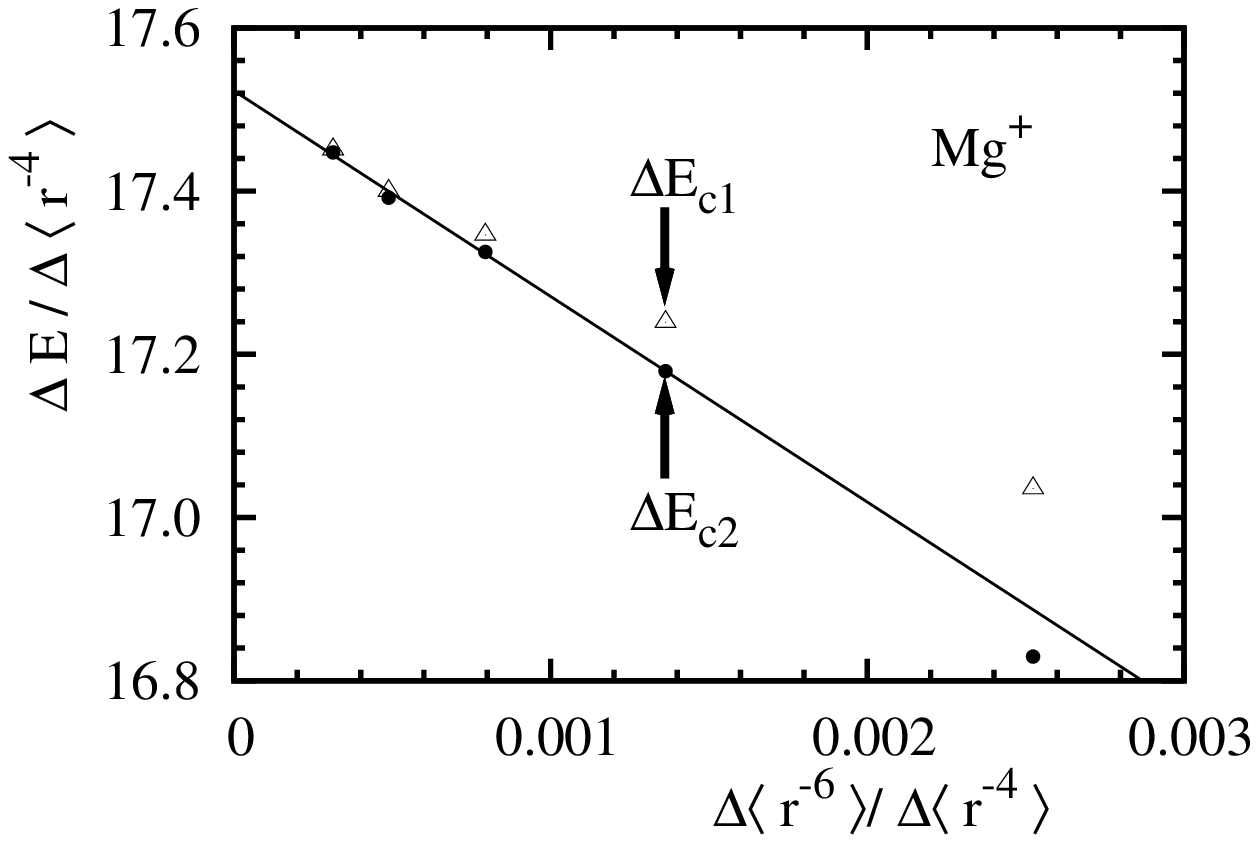}
\vspace{0.02cm}
\caption[]{ \label{mg+}
The polarization plot of the fine-structure intervals of Mg for the $n = 17$ 
Rydberg levels.  The $\Delta E_{c1}$ intervals are corrected for relativistic, 
second-order and Stark shifts.  The $\Delta E_{c2}$ intervals account for 
$\langle r^{-7} \rangle$ and $\langle r^{-8} \rangle$ shifts. 
The linear regression for the $\Delta E_{\rm c2}$ plot did not
include the last point.  
}
\end{figure}

\begin{figure} [th]
\includegraphics[width=8.50cm,angle=0]{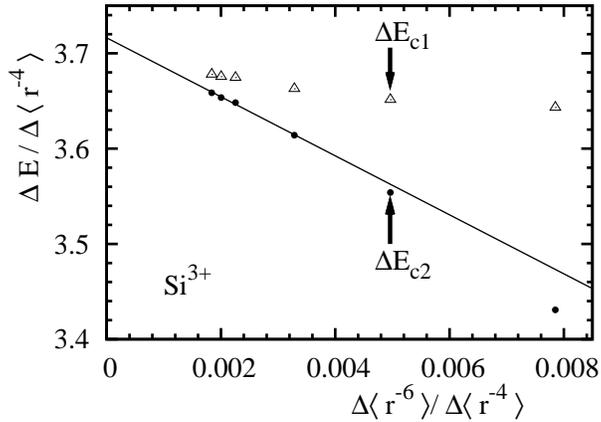}
\vspace{0.02cm}
\caption[]{ \label{si3+}
The polarization plot of the fine-structure intervals of Si$^{2+}$ for the 
$n = 29$ Rydberg levels.  The $\Delta E_{c1}$ intervals are corrected for 
relativistic, second-order and Stark shifts.  The $\Delta E_{c2}$ intervals 
account for $\langle r^{-7} \rangle$ and $\langle r^{-8} \rangle$  
shifts.  The linear regression for the $\Delta E_{\rm c2}$ plot did not
include the last two points.  
}
\end{figure}

\subsection{Si$^{3+}$}

The polarization plot for Si$^{3+}$ is shown in Figure \ref{si3+}. 
The most notable feature is the large difference between the
$\Delta E_{\rm c1}$ and $\Delta E_{\rm c2}$ data-sets.  The other
notable feature is the pronounced deviation from linear of the 
$\frac{\Delta E_{c2}}{\Delta \langle r^{-4} \rangle}$ plot.     

Examination of Table \ref{DeltaE} for the (29,8)-(29,9) interval shows 
that the net $\Delta E_{7,8,8L}$ correction is very close in magnitude to 
the $\Delta E_6$ energy correction.  The $\Delta E_{7,8,8L}$ correction 
is still more than 50$\%$ of the $\Delta E_6$ correction for the 
(29,10)-(29,11) interval.  The polarization series is an 
asymptotic series \cite{dalgarno56a,drachman82a} and is not absolutely 
convergent as 
$n$ increases.  As mentioned by Drachman
\cite{drachman82a}, a condition for the usefulness of the polarization
series is that the $\Delta E_{7,8,8L}$ corrections should be 
significantly smaller than the $\Delta E_6$ corrections.  This condition 
is not satisfied for the first two intervals and leads to the noticeable 
curvature in the plot of the $\Delta E_{\rm c2}$ data points.

The resolution to this problem would be to increase the $L$ values at which 
the intervals are measured. But Stark shift corrections become increasingly 
important at high $L$.  The Stark shift corrections are significant for the 
(29,11)-(29,14) interval.

A line of best fit was drawn using the four data points with 
$\Delta \langle r^{-6} \rangle /\Delta \langle r^{-4} \rangle < 0.004$.
The linear regression gave an intercept of $B_4 = 3.7163(32)$ and a slope 
of $B_6 = -30.96(134)$.  The intercept translates to a polarizability of 
7.433 a.u..  To put this in perspective, the polarizability originally 
deduced from the RESIS experiment was 7.408(11) \cite{komara03a}.  
A later analysis which included the $C_7$ and $C_{8L}$ potentials gave 
7.426(12) a.u. \cite{snow07a}.  There has been a steady increase in 
the derived dipole polarizability as more higher-order terms in the 
polarization series are incorporated into the analysis.   

The polarization plot $B_6$ of $-$30.96(134) was about 10$\%$ larger in 
magnitude than the MBPT-SD value of $-$27.06(5).  This value  
of $B_6$ results in a quadrupole polarizability of 
$\alpha_2 = (-2 \times 30.96 + 6 \times 11.04) = 4.34$ a.u. which
is 60$\%$ smaller than the HFCP and MBPT-SD polarizabilities.  The uncertainty 
of $(2 \times 1.34 + 6 \times 0.006) =  3.0$ a.u. is too small 
to allow consistency with the theoretical values.   

\subsection{An alternate perspective}

The analysis so far can be regarded as a standard polarization
analysis but with additional refinements due to improved 
knowledge about the higher-order terms in the polarization
series.  However, it is worthwhile to examine the analysis 
from a different perspective.  

The comparison between first principles theory and the RESIS 
experiment has resulted in agreement to better than 1$\%$ for 
dipole polarizabilities. The quality of the agreement for the 
quadrupole polarizability is not nearly so good.  But can 
the analysis of the RESIS experiment be expected to
yield quadrupole polarizabilities that are a serious test 
of calculation?  The quadrupole polarizability is derived from 
the slope of the polarization plot.  But the higher-order
polarization corrections and Stark shifts result in energy 
corrections that amount to between 30-100$\%$ of the raw 
$C_6$ energy shift.  And it must be recalled that the 
polarization series itself is an asymptotic series 
\cite{dalgarno56a} so there are uncertainties 
about the size of omitted terms.  

One way forward is to use the dipole polarizability comparison   
as a guide to the accuracy of the quadrupole polarizability.  
The first principles dipole polarizabilities are expected to be 
accurate to better than 0.5$\%$ and this has been confirmed by 
experiment.  As discussed earlier, the uncertainty in the quadrupole 
polarizability for Na-like ions should be smaller than the dipole 
polarizability.  Therefore it is not credible 
to postulate large errors in the atomic structure calculations 
of the quadrupole polarizability on the basis of a $B_6$ derived 
from the polarization plot.  It makes more sense to use the 
theoretical $C_6$ to estimate the size of unaccounted systematic 
effects in the measured energy shifts.   

The large uncertainties in $B_6$ do not detract greatly from 
the the accuracy of the dipole polarizability.  One of the 
reasons higher-order effects can substantially impact $B_6$ 
is that $\Delta E_6$ is small because of the cancellation 
between $\alpha_2$ and $\beta_1$.  However, the relatively 
small size of $B_6$ means a large uncertainty in $B_6$ 
has a relatively small impact on the derived $\alpha_1$.    

The impact of possible systematic errors on Mg$^+$ was determined 
by redoing the linear regression with a fixed value of $B_6$ that 
was constrained to lie between $-251.2 \pm 19.0$.  The uncertainty 
of 19.0 was derived by adding the statistical uncertainty of 7.9 
from the initial linear regression fit to $|240.1-251.2|$, the 
difference between the $C_6$ of Table \ref{Cnvalues} and the 
initial $B_6$ from the linear regression.  This gave a 
revised uncertainty of $\delta B_4 = 0.015$, leading to a final 
$\alpha_1$ of 35.04(3) a.u..    

The same analysis was repeated for Si$^{3+}$.  In this instance 
the derived value of $\alpha_1$ was 7.433(25) a.u..   

\subsection{Estimate of the resonant oscillator strengths.}

As the polarizabilities are dominated by the resonant transition 
it is possible to derive an estimate for the resonant multiplet  
strength \cite{safronova04b}.  We use the relation  
\begin{equation} 
S_{3s  -  3p} =  \frac{  \alpha_1 - \alpha'_1  - \alpha_{\rm core}  } 
{ \displaystyle{ \frac{2}{9\Delta E_{3s\!-\!3p_{1/2}}} + \frac{4}{9\Delta E_{3s\!-\!3p_{3/2}}}   } }  \ .  
\label{alphaS} 
\end{equation} 
In this expression $\alpha_1$ is the polarizability extracted from the 
polarization plot while $\alpha_{\rm core}$ is the net core polarizability, 
and $\alpha'_1$ is the valence polarizability excluding the resonant 
transition.  For the Mg$^+$ multiplet, we use  
\begin{equation} 
S_{3s-3p} = \frac{ 35.044 - 0.112  - 0.463}{4.08436} =  8.439  \ .  
\label{SMg+} 
\end{equation} 
Using the uncertainties detailed earlier, the final value is  
8.439(11).  This is equivalent to a line strength of 
$S_{3s-3p_{3/2}} = 11.25(2)$, in agreement with the recent
experimental value of 11.24(6) \cite{hermann08a}.    

Repeating the analysis for Si$^{3+}$ gave a multiplet strength of
3.519(16) for the $3s \to 3p$ transition. This is equivalent to
$S_{3s-3p_{3/2}} = 4.693(24)$ which is 0.14$\%$ larger than
the MBPT-SD line strength of 4.686.    
 
\section{Conclusions}

A survey of polarization parameters of the Mg$^+$ and Si$^{3+}$ ion states 
relevant to the analysis of the RESIS experiments by the Lundeen group 
\cite{komara03a,snow07a,snow08a} have been presented by two complementary 
approaches.  The reanalysis of the fine-structure intervals gave dipole 
polarizabilities of 35.04(4) a.u. for Mg$^+$ and 7.433(25) a.u. for
Si$^{3+}$.  The HFCP and MBPT-SD calculations give polarizabilities that 
lie within 0.2$\%$ of each other for Mg$^+$ and 0.3$\%$ for Si$^{3+}$.  The 
{\em ab-initio} MBPT-SD dipole polarizabilities of 35.05(12) and 7.419(16) 
a.u. respectively agree with the experimental dipole polarizabilities to 
accuracy of better than 0.3$\%$.    

One notable feature of the present analysis is the very good agreement 
between the HFCP and MBPT-SD calculations.  Indeed, the MBPT-SD calculation agrees  
better with the computationally simple HFCP calculation, than it does 
with two very large CI type calculations.  For example, the polarizabilities 
of the completely ab-initio CI calculation \cite{hamonou07a} are about 
1.5$\%$ larger than the MBPT-SD and HFCP  polarizabilities.   We conclude that 
a semi-empirical calculation based on a HF core can easily be superior to 
a pure CI calculation unless the CI calculation is of very large dimension.  
The HFCP approach has the advantage of tuning the model energy levels to 
experiment and this goes a long way to ensuring that many of the 
interesting observables will be predicted accurately.  There is one 
feature common to the HFCP and MBPT-SD approaches.  Both approaches 
approximate the physics of the dynamical corrections beyond HF/DF, but 
within those approximations an effectively exact calculation is made.    

There are two major sources of systematic error that can impact the
interpretation of the RESIS experiment.  To a certain extent one has 
to choose the ($n$,$L$) states to navigate between the Scylla \cite{homerBC} 
of non-adiabatic corrections and the Charybdis \cite{homerBC} of Stark 
shifts.  If $L$ is too small, then the $\Delta E_{7,8,8L}$ shift becomes 
larger than $\Delta E_{6}$, thus invalidating the use of Eq.~(\ref{vpol}).  
On the other hand, Stark shift corrections become increasingly bigger 
as $L$ becomes larger.  These problems are most severe in Si$^{3+}$ and 
are responsible for the slope of the polarization curve being different 
from the atomic structure predictions.  An explicit two-state model of 
long range polarization interactions is probably needed to realize the 
full potential of the RESIS experiment   

\begin{acknowledgments}

This work was supported in part by the National Science Foundation 
Grant No.\ PHY-0758088.

\end{acknowledgments}


\end{document}